\documentclass[aps,prl,twocolumn,showpacs,superscriptaddress]{revtex4}
\usepackage{epsfig}
\def\avg#1{\langle#1\rangle}
\def\Re{\mbox{Re}}
\def\Im{\mbox{Im}}
\def\be{\begin{equation}}       \def\ee{\end{equation}}
\def\bea{\begin{eqnarray}}      \def\eea{\end{eqnarray}}
\def\PRB{Phys. Rev. B}
\def\PRL{Phys. Rev. Lett.}

\begin{document}
\title{Spin-orbit coupling induced magnetism in the $d$-density wave phase
of La$_{2-x}$Ba$_x$CuO$_4$ superconductors
}
\author{Congjun Wu}
\affiliation{Department of Physics, McCullough Building,
Stanford University, Stanford CA 94305-4045}
\affiliation{Kavli Institute for Theoretical Physics, University
of California, Santa  Barbara, CA 93106}
\author{Jan Zaanen\footnote{On leave of absence from the Instituut-Lorentz
for Theoretical Physics, Leiden University, Leiden, The Netherlands.}
}
\affiliation{Department of Physics, McCullough Building,
Stanford University, Stanford CA 94305-4045}
\author{Shou-Cheng Zhang}
\affiliation{Department of Physics, McCullough Building,
Stanford University, Stanford CA 94305-4045}

\begin{abstract}
We study the effects of spin-orbit coupling in the $d$-density
wave (DDW) phase. In the low-temperature orthorhombic phase of
La$_{2-x}$Ba$_x$CuO$_4$, we find that spin-orbit coupling induces
ferromagnetic moments in the DDW  phase, which are polarized along
the [110] direction with a considerable magnitude. This effect
does not exist in the superconducting phase. On the other hand, if
the $d$-density wave order does not exist at zero field, a
magnetic field along the [110] direction always induces such a
staggered orbital current. We discuss experimental constraints on
the DDW states in light of our theoretical predictions. 
\end{abstract}
\maketitle

The mechanism of the pseudogap phenomena in high $T_c$
superconductors remains controversial. Chakravarty {\it et al.}
\cite{chakravarty2001} proposed that it may originate from a
hidden long-range  $d$-density wave (DDW) order \cite{nayak2000,affleck1988},
which competes with the $d$-wave superconductivity (DSC). 
This scenario has aroused much interest. Extensive analytic
and numerical investigations have shown its existence under
certain conditions in a variety of  one  and two dimensional
systems \cite{fjarestad2002,wu2003,tsuchiizu2002,schollwock2003,capponi2004}.
However, these states are hard to detect experimentally and results are
still controversial \cite{chakravarty2001a}. 
Polarized neutron scattering experiments \cite{mook2002,mook2004} in
YBa$_2$CuO$_{6+x}$
show some supporting evidence. On the other hand, Stock et al.
\cite{stock2002} found no indication
of this phase using non-polarized neutron beams.      

Recently, spin-orbit (SO) coupling has received 
much attention in the emerging science of spintronics.
Murakami, Nagaosa and Zhang proposed the intrinsic spin Hall effect
through SO coupling in the $p$-doped semiconductors to generate
the dissipationless spin current using electrical fields  \cite{murakami2003}.
Similar effects were also predicted in the $n$-doped systems \cite{sinova2004}.
The spin-Hall effect in GaAs  has already been observed experimentally
\cite{spinhall}.

In Mott-insulators, SO coupling also has important effects
on the Heisenberg superexchange interactions,
which is responsible for the anisotropic correction termed
the Dzyaloshinskii-Moriya (DM) interaction \cite{dzyaloshinskii-moriya}.
Given the intrinsic spin Hall effect in the semiconductors,
it is natural to ask what happens in the presence of
SO coupling with states carrying spontaneous
electrical currents  like the DDW state. The answer turns
out to depend on details,  but these conspire in the La$_2$CuO$_4$
system to give rise to an experimentally observable effect: as in the
half-filled antiferromagnets, SO coupling gives rise to a
weak planar ferromagnetism which can be used to detect
this otherwise elusive phenomenon.

At zero temperature, La$_{2-x}$Ba$_x$CuO$_4$ undergoes a structural
phase transition  from the low-temperature 
orthorhombic (LTO) phase to the low temperature tetragonal (LTT) 
phase at doping $\delta_c\approx 0.12$.
In the LTO phase of the undoped La$_2$CuO$_4$, 
the DM interaction originates from the staggered distortion pattern of
the oxygen octahedrals.
It results in the antiferromagnetic moments
lying in the $ab$-plane and the weak ferromagnetic moments along
the $c$-axis \cite{thio1988}.
By analogy, the DDW state exhibits staggered orbital moments.
Without SO coupling, the DDW state decouples from the spin channel,
which  remains paramagnetic.
However, SO coupling couples these two channels together,
and thus orbital currents should affect spin
and lead to  observable effects.

In this article, we find that the staggered orbital current induces 
uniform ferromagnetic moments in both LTO and LTT phases.
The moments lie in the [110] direction in the LTO phase and
the [100] direction in the LTT phase, respectively.
The magnitude per Cu cite is at the order of several percent of one
Bohr magneton, thus is detectable.
Conversely, if the DDW order does not exist in the ground state,
it can be induced by magnetic fields,
suggesting that the degree of proximity to the instability can
be in principle investigated as well.

\begin{figure}
\centering\epsfig{file=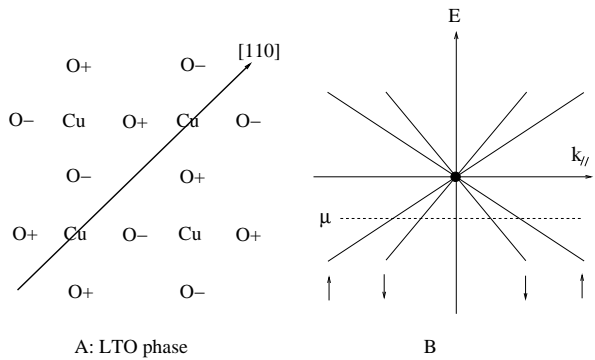,clip=1,width=0.9\linewidth,
angle=0}
\caption{
A) The lattice symmetry in the LTO phase. $O+$ ($O-$) denotes
the oxygen atom moving into (out of)  the CuO plane.
The arrow indicates the [110] direction.
B) The anisotropic Dirac cone-like dispersion relations
around nodes $(\pm \pi/2,\pm \pi/2)$ for spin parallel ($\uparrow$)
and anti-parallel ($\downarrow$) to the [110] direction.
}
\label{dispersion}
\end{figure}

We consider the mean field DDW Hamiltonian with SO coupling 
in the LTO phase:
\begin{eqnarray}
H_{MF}&=&\sum_{\avg{ij}\alpha\beta} \Big \{ c^\dagger_{i\alpha}
(-t_{\rm{eff}}+ i\vec \lambda_{ij} \cdot \vec \sigma_{\alpha\beta})
c_{j\beta}+h.c. \Big\} \nonumber \\
&+& i \sum_{\avg{ij}\sigma}  \Im\chi_{ij}
(c^\dagger_{i\alpha}c_{j\alpha}-h.c.)
-\mu\sum_{i\sigma} c^\dagger_{i\sigma} c_{i\sigma} \nonumber\\
&+&\frac{1}{2V} \sum_{\avg{ij}} {\Im} \chi_{ij} \Im\chi_{ij},
\label{eq:ham}
\end{eqnarray}
where $\avg{ij}$ indicates summation over the nearest neighbors only.
$\chi_{ij}$ is the decoupling of the Heisenberg exchange
term in the particle-hole channel \cite{affleck1988}
as
$\chi_{ij} = V\avg{c^\dag_{i\sigma}
c_{j\sigma}}$.
Its imaginary part $\Im \chi_{ij}$ is the DDW order parameter
and is treated self-consistently below.
On the other hand, $\Re \chi_{ij}$ changes slowly 
within the parameter regime discussed below, thus
is absorbed into the effective hopping integral $t_{eff}$
which gives the band width of holes.
We choose $t_{eff} \approx 100$meV
which is rescaled to 1 below.
The SO coupling term $\vec \lambda_{ij}$  is determined by
the lattice symmetries  \cite{coffey1991} as shown in
Fig. \ref{dispersion}. A, such as
i) two-fold rotations around the $c$-axes passing the in-plane O sites,
ii) inversion symmetry with respect to Cu sites,
and iii) reflection symmetry with  respect to the [110] direction.
Consequently, it shows a  staggered pattern 
$
\vec \lambda_{i,i+\hat x}=(-)^{i_x+i_y} (\lambda_1,\lambda_2,0),
\ \ \
\vec \lambda_{i,i+\hat y}=(-)^{i_x+i_y} (-\lambda_2,-\lambda_1,0)
$ \cite{bonesteel1992}.
$\vec \lambda_{ij}$ is almost 
perpendicular to the bond direction in the LTO phase
\cite{shekhtman1992}, i.e., 
$\lambda_1\ll\lambda_{2}$,
and  $\lambda_2$ is estimated  to be around 2$m$eV
\cite{coffey1991}.
In the LTO phase, the symmetries
ii) and iii) still ensure the  $\Im \chi_{ij}$  to exhibit
the $d_{x^2-y^2}$ pattern.

We first present physical arguments for the appearance of 
the ferromagnetic moments for a simplified case of $\lambda_1=\lambda_2
= \lambda/\sqrt 2$, and then show that the realistic values of $\lambda_{1,2}$
essentially give the same result.
In the simplified case, the SO coupling term plays the role of the 
staggered spin flux with the quantization axis along the [110]
direction \cite{bonesteel1992}.
The effective 
Hamiltonian reads
\bea
H&=&\sum_k c^\dagger_{k\alpha} c_{k\alpha} (\epsilon_k-\mu )
+2 i  f(\vec k) c^\dagger_{k\alpha}
\big\{  \lambda \sigma_{1,\alpha\beta} \nonumber \\
&+&\Im\chi  \delta_{\alpha\beta}
\big \} c_{k+Q,\beta},
\label{eq:ddw}
\eea
where $f(\vec k)=\cos k_x-\cos k_y$, $\sigma_1= (\sigma_x+\sigma_y)
/\sqrt 2$, $Q=(\pi,\pi)$, and the spin index $\alpha=\uparrow (\downarrow)$ 
refers to parallel (anti-parallel) to the [110] direction.
For $\vec k$ around the nodes $(\pm\pi/2,\pm\pi/2)$, we define
$\vec k_\parallel$ and $\vec k_\perp$ to be the projections of its
deviation from the nodes on the directions
parallel and perpendicular to the nested Fermi
surface, respectively.
As shown in Fig. \ref{dispersion} B, we obtain the anisotropic Dirac-cone
like dispersion relation with different slopes for spin $\uparrow(\downarrow)$
electrons
\bea
E(k)_{a,\alpha}= \pm \sqrt{v^2_f k^2_\perp +8({\Im \chi}
\pm   \lambda)^2 k_\parallel^2},
\label{eq:dispersion}
\eea
where the first $\pm$ on the RHS of Eq. (\ref{eq:dispersion})
corresponds to the upper (lower) Dirac cone as denoted by
the band index $a$,
and the second $\pm$ corresponds to the spin directions
$\uparrow$ ($\downarrow$), respectively.
At half-filling, no ferromagnetic moments exist because the lower two bands
with opposite spin configurations are both fully occupied.
However, at finite doping $\delta$, they are  occupied
differently,
thus a spin polarization appears along the [110] direction.
At very small doping, the ferromagnetic moment per site can be estimated
from the dispersion relation of  Eq. (\ref{eq:dispersion}) as
\bea
\frac{M}{\delta}= \left.\Big\{
\begin{array}{l}
| \lambda/\chi| \ \ \ \mbox{at} \ \ \  |\lambda/\chi|\ll 1 \\
|\chi/\lambda|  \ \ \ \mbox{at} \  \ \ |\chi/\lambda|\ll 1 \\
\end{array} \right. . \label{polar}
\eea

\begin{figure}
\centering\epsfig{file=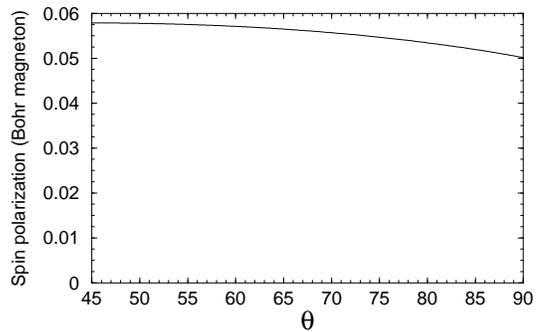,clip=1,width=70mm, angle=0}
\caption{Spin polarization in the LTO phase at doping $\delta=0.1$
with  $\lambda_1 =\lambda \cos \theta, \lambda_2=\lambda
\sin\theta$ ($45^\circ\le\theta\le 90^\circ$ ).
The results within the range $0^\circ\le\theta\le 45^\circ$
are symmetric to the case of  $90^\circ-\theta$. }
\label{angle}
\end{figure}

\begin{figure}
\centering\epsfig{file=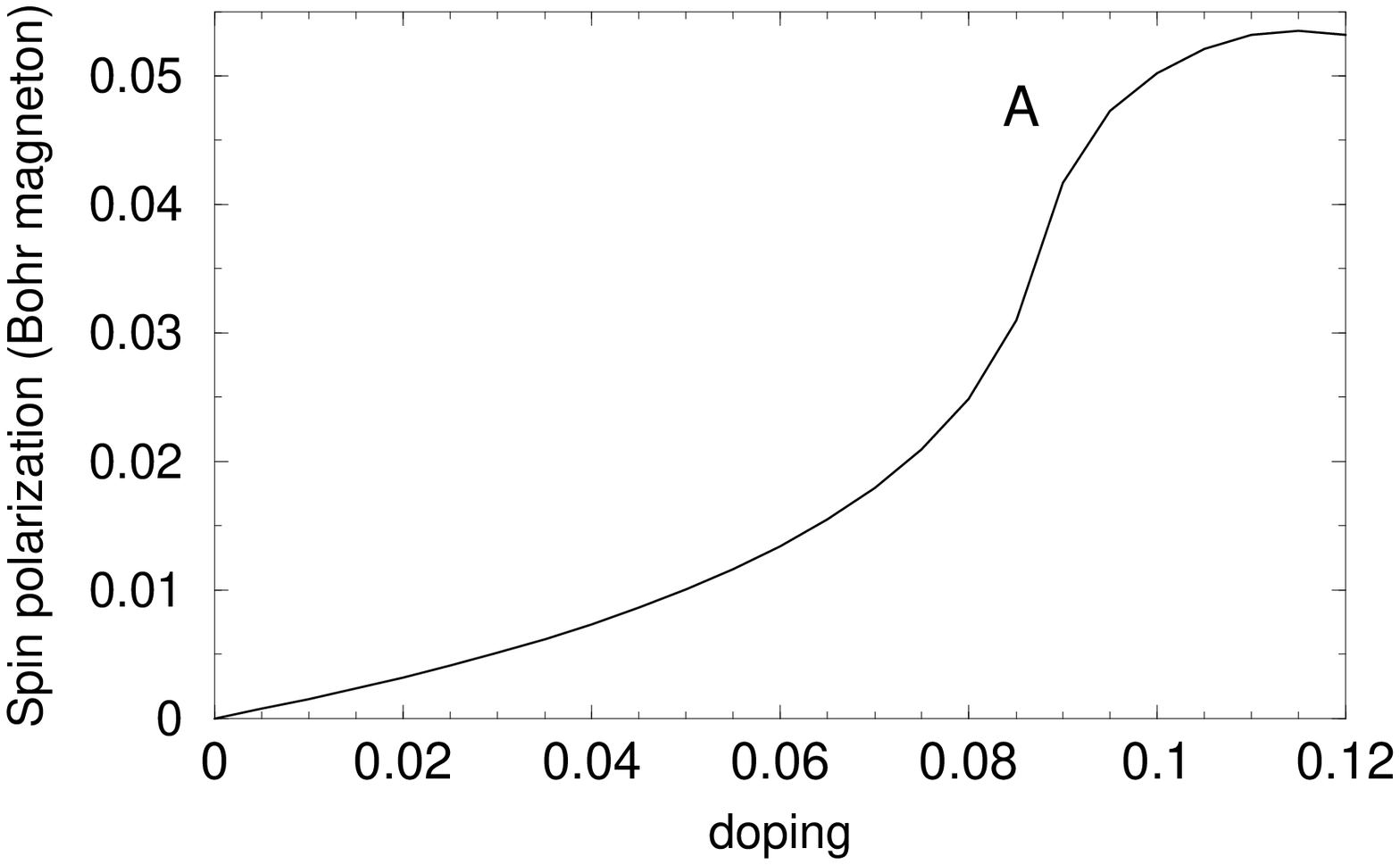,clip=1,width=70mm, angle=0}
\\
\centering\epsfig{file=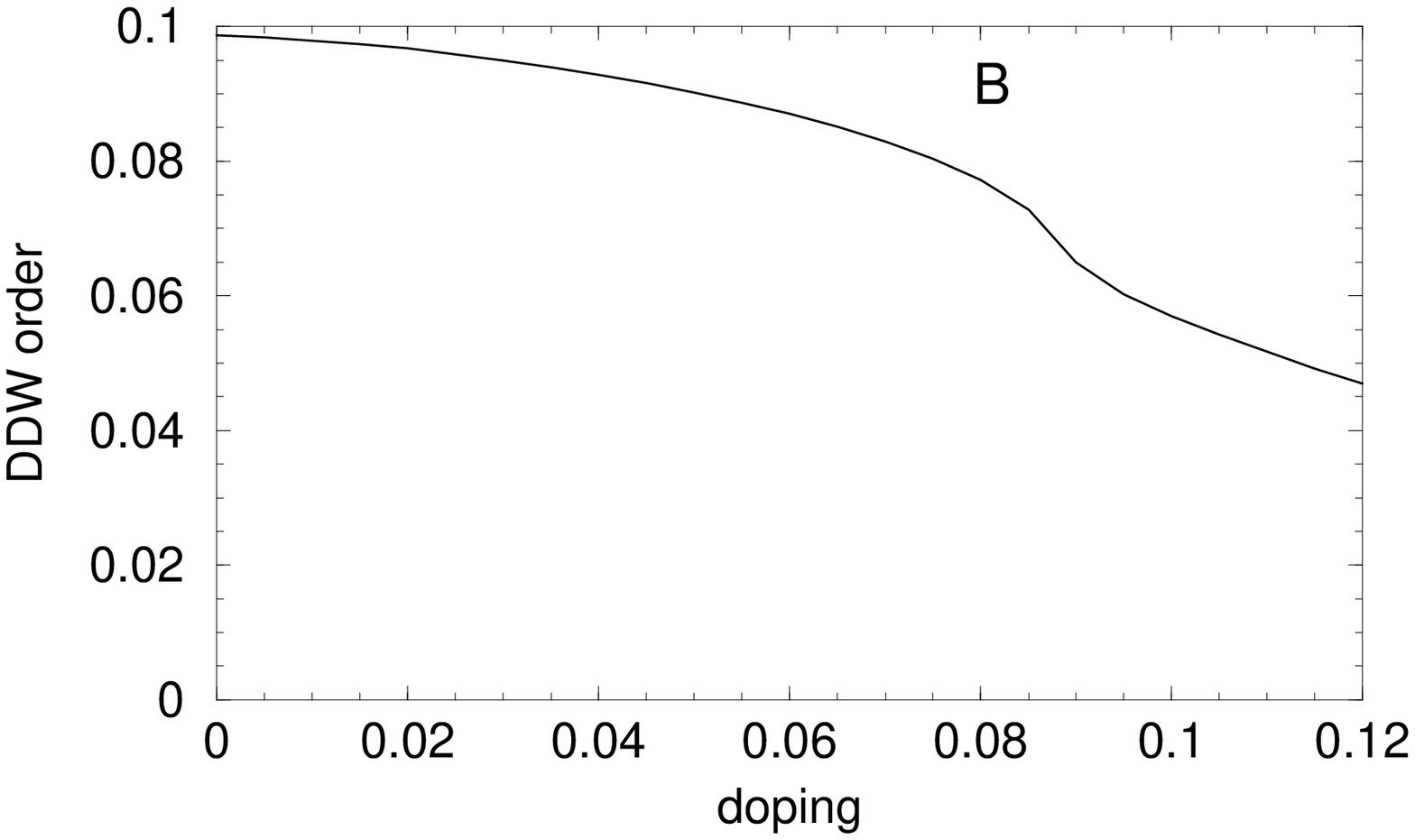,clip=1,width=0.8\linewidth, angle=0}
\caption{ A) Spin polarizations $M$ v.s. doping levels $\delta$ at T=0K.
B) The DDW order parameter $\Im \chi$ v.s. doping levels $\delta$ at T=0K.}
\label{selfcon}
\end{figure}

It is instructive to consider the underlying symmetry reasons
for this effect: the DDW order breaks the time reversal (TR)
symmetry in the orbital channel
while ferromagnetism breaks it in the spin channel.
SO coupling couples two channels together to linear order.
We emphasize that the polarization along the [110] direction
is valid for {\it  general} values of $\lambda_{1,2}$.
This is  protected by the following symmetry structures.
Although the symmetries  i) and iii) are broken by the DDW order,
their combination with the TR operation together still leave
the system invariant.
These symmetries fix the only possible spin polarization along the [110]
direction, and further exclude the antiferromagnetic order.

Now we consider the general values of $\lambda_{1,2}$ with the 
parameterization  $\lambda_1=\lambda \cos \theta, \lambda_2=\lambda
\sin \theta~ (\theta=0^\circ \sim 90^\circ )$.
The realistic values in La$_{2-x}$Ba$_x$CuO$_4$, i.e., 
$\lambda_1\ll \lambda_2$, corresponds to $\theta=90^\circ$.
The new effective Hamiltonian includes Eq. (\ref{eq:ddw}) 
but with the replacement of $\lambda$ with $\lambda 
(\cos \theta +\sin \theta)/\sqrt 2$, 
and also an extra term of
\bea
\Delta H=-2 i\sum_{\vec k,\alpha\beta}\Big\{
\Delta  \lambda ~ g(k)
c^\dagger_{k\alpha} \sigma_{2,\alpha\beta}
c_{\vec k+\vec Q,\beta}-h.c.\Big\},
\eea
where $g(k)=\cos k_x +\cos k_y$,
$\sigma_2= (\sigma_x-\sigma_y)/\sqrt 2$,
and $\Delta \lambda= \lambda (\sin \theta-\cos\theta)/\sqrt 2$.
Correspondingly, the spin quantization axis
of the electron eigenstate depends on the momentum $\vec k$,
which deviates from the [110] direction with an
angle $\phi_{k}$ satisfying
$\tan \phi_k= \frac{k_\perp}{k_\parallel} \frac{\sin \theta-\cos \theta}
{\sin \theta +\cos \theta} $.
This helical structure reduces the magnitude of the ferromagnetic
moments.
However, due to the fact that $t_{eff}$ is much larger than
$\chi$ and $\lambda$, the Dirac-cone is highly  anisotropic.
Then  $k_\parallel \gg k_\perp$ holds on  most part of the Fermi pocket, 
and the spin deviates from the [110] direction only at small angles.
In other words, 
the induced ferromagnetic moment is insensitive to the
ratio of $\lambda_{1}/\lambda_2$.

In Fig \ref{angle}, we show the numerical results for the spin polarization 
per site $M$ with the general values of $\theta$ at $\delta=0.1$ 
and $T=0K$ by using the standard self-consistent method \cite{wu2002}.
We choose parameters $\lambda=0.02, V=0.22$ to agree with the physical
value of SO coupling and to arrive at a reasonable pseudogap energy scale
in La$_{2-x}$Ba$_x$CuO$_4$.
In the realistic case of $\theta=90^\circ$, the polarization only
decreases about $15\%$ compared with its maximal value at $\theta=0^\circ$.
For all numeric results below, we keep $\theta=90^\circ$,
i.e., $\lambda_1=0$.
We further show $M$ and $\Im \chi$ v.s. the doping $\delta$ at $T=0K$
in Fig. \ref{selfcon}A and
\ref{selfcon}B.
At low doping $\delta$, $M$ indeed scales with $\delta$
linearly as indicated in Eq. (\ref{polar}).
As $\delta$ increases, $\Im\chi$ drops, and consequently $M$
increases faster than linearly.

\begin{figure}
\centering\epsfig{file=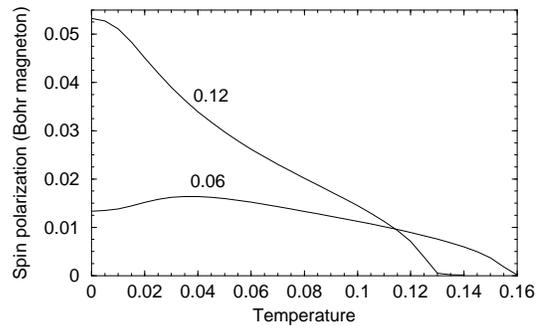,clip=1,width=70mm, angle=0}
\caption{Spin polarization v.s. temperatures at $\delta=0.06,0.12$.
Temperatures are in the unit of $t_{eff}$.
}
\label{Temp}
\end{figure}

\begin{figure}
\centering\epsfig{file=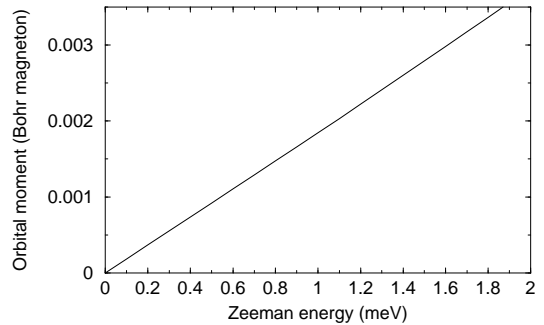,clip=1,width=70mm, angle=0}
\caption{Staggered orbital moment in the unit of $\mu_B$ v.s.
the Zeeman energy $E_{z}= g/2 B \mu_B $, where $t_{eff}=100$meV,
$\lambda/t_{eff}=0.02$, $\Delta \lambda=0$, and
the Lande factor $g=2$ .
$\vec B$ is along the [110] direction.
}
\label{fig:orbit}
\end{figure}

The finite temperature behavior of  the induced polarization $M$
is also interesting as shown in Fig. \ref{Temp}.
At small dopings (e.g. $\delta=0.06$), $M$  increases slowly at low
$T$ and decreases after $T$ passes an intermediate value,
while $M$ decreases monotonically at high dopings (e.g. $\delta=0.12$).
At low dopings, $\Im\chi$ is large at $T=0K$ which decreases with
increasing $T$.
As a result, the ratio $\lambda/\Im\chi$ increases, and so does $M$
as indicated in Eq. (\ref{polar}).
As $T$ goes large, then the thermal effect dominates and
$M$ decreases.
At high dopings where $\Im\chi$ is comparable to $\lambda$,
$M$ depends less sensitively on  $\lambda/ \Im\chi$,
and thus thermal effects dominates in the whole temperature range.

The magnitude of the induced moments is at the order of $10^{-2} 
\mu_B$ which should be detectable. Due to their 
ferromagnetic nature, domain wall structures should be formed
and no macroscopic magnetic field is present.
A hysteresis behavior appears when an external in-plane magnetic
field is applied to the sample. In the neutron scattering 
experiment, the elastic Bragg peaks at reciprocal lattice vectors are 
the evidence for these moments. The muon spin relaxation is also sensitive to
the internal magnetic fields. To our knowledge, no such  effects have 
been detected in the LTO phase of La$_{2-x}$Ba$_x$CuO$_4$.

SO coupling also leads to  a ``staggered spin galvanic effect''
as the inverse of the DDW induced ferromagnetism.
Assuming that the DDW order does not exist in the ground state,
or equivalently setting  $V=0$ in Eq. (\ref{eq:ham}), and adding the
Zeeman energy term $H_{z}= -\sum_i g\mu_B\vec B \cdot \vec S_i$,
we find that a spin polarization along the [110] direction induces
a staggered orbital current.
The staggered current per bond is written as
\bea
I_{stag}&=&\frac{i e t_{eff}}
{N\hbar}\sum_{k\sigma} c^\dagger_{k\sigma} c_{k+Q\sigma} f(\vec k)
+\frac{e}{\sqrt 2 N \hbar }\sum_{k,\alpha\beta} c^\dagger_{k\alpha}
c_{k\beta} 
\nonumber \\
&\times&\Big\{(\lambda_1 \cos k_x
+\lambda_2 \cos k_y) \sigma_{x,\alpha\beta} \nonumber \\
&+&(\lambda_2 \cos k_x +\lambda_1 \cos k_y) \sigma_{y,\alpha\beta}
 \Big\},
\eea
where the second term originates from SO coupling.
Under the symmetry analysis,
$I_{stag}$ can be induced only by the $B$ field 
along the [110] direction.
Using the values of $t_{eff}$ and $\lambda$ stated above,
the lattice constant $a=3.8$ \AA, and  $\theta=90^\circ$,
we show in Fig. \ref{fig:orbit} the linear behavior of the staggered orbital
moment per plaquette v.s. Zeeman energy $E_z=\frac{1}{2} g \mu_B B$.
The magnitude reaches
the order of  $10^{-3}$ $\mu_B$ at $E_z\approx 1$meV
which corresponds to $B \approx 10 T$.
The typical value of the DDW orbital moment estimated theoretically
is at the order of $10^{-2} \mu_B$ \cite{chakravarty2001a}.
Compared with it, our induced orbital moment is about one order
smaller.

In the LTT phase, the lattice symmetry results in a different staggered
SO coupling pattern  as
$ \vec \lambda_{i,i+\hat x}=(-)^{i_x+i_y} (\lambda_1,0,0),
\vec \lambda_{i,i+\hat y}=-(-)^{i_x+i_y} (\lambda_2,0,0)
$ \cite{bonesteel1992},
where the spin quantization is fixed along the [100] direction
and also $\lambda_1 \ll \lambda_2$ \cite{shekhtman1992}.
Similar analysis indicates that the previous results 
also apply here with the replacement of the [110]
with the [100] direction.

Next we discuss the effect of SO coupling in the superconducting
portion of the  phase diagram.
SO coupling does not change the nature of the DSC phase in 
the absence of the DDW order.
Due to the conservation of the TR symmetry, no ferromagnetic moment 
can appear. Although the spin and the 4-fold rotational symmetries are broken, 
symmetries ii) and iii) still leave the $d$-wave singlet pairing
structure unchanged, and thus the node quasiparticles are kept.
In the coexisting region of DDW and DSC, if the DDW order is large
compared with the DSC, the induced ferromagnetism
suppresses the superconductivity, which may lead to
Fulde-Ferrell-Larkin-Ovchinnikov phases \cite{FFLO}.  

At last, we briefly discuss  the YBa$_2$Cu$_3$O$_{6+\delta}$
system.
The inversion symmetry is broken in  each CuO plane, and the
resulting SO coupling is uniform in contrast to the staggered pattern
in La$_{2-x}$ Ba$_x$CuO$_4$ systems.
Due to the existence of CuO chains, the [100] and [010] directions are
not equivalent any more, and the 4-fold rotational symmetry is broken,
while the reflection symmetries respect to [100] and [010] directions
are still kept.
A straightforward symmetry analysis gives the form of  the SO coupling
\cite{thio1988} as
$\vec \lambda_{i,i+\hat x}= (0, \lambda_1,0), ~~
\vec \lambda_{i,i+\hat y}= (\lambda_2,0,0).
$
However, because the staggered orbital current still preserves the
above two reflection symmetries, this SO coupling can not induce magnetic
moments on Cu sites. We notice that at least in a single plane
the broken parity
has a significant effect in the superconducting phase: as in the
2D Rashba system the singlet and triplet pairing channels should be
mixed \cite{gorkov2001}.

In summary, we have investigated the effect of SO coupling
to the DDW state in doped La$_{2-x}$Ca$_x$CuO$_4$.
If the DDW state indeed exists, SO coupling results in a uniform
ferromagnetic moment along the [110] direction in the LTO phase or
the [100] direction in the LTT phase.
This effect can be used to test the validity of the DDW scenario
for the pseudogap mechanism.
The inverse effect is also predicted that a in-plane Zeeman field
induces a staggered orbit moment.

We thank S. Chakravarty and O. Vafek for helpful discussions.
This work is supported by the NSF under Grant No. DMR-0342832,
and U.S. Department of Energy, Office of Basic Energy Sciences
under Contract No. DE-AC03-76SF00515.
CW is supported by the Stanford Graduate Fellowship,
and the NSF Grant No. Phy99-07947.
JZ acknowledges financial support by the Fulbright
foundation in the form of a senior fellowship.


\begin{thebibliography}{10}

\bibitem{chakravarty2001}
S. Chakravarty et al., Phys. Rev. B {\bf 63}, 094503-1 (2001).

\bibitem{nayak2000}
C. Nayak,  Phys. Rev. B {\bf 62}, 4880 (2000).

\bibitem{affleck1988} I. Affleck, {\it et. al.}, Phys. Rev. B {\bf
38}, 745 (1988).


\bibitem{fjarestad2002} J. O. Fjaerestad {\it et. al.}
\PRB~ {\bf 65}, 125106 (2002).

\bibitem{wu2003} C. Wu {\it et. al.}, Phys. Rev. B {\bf 68},
115104 (2003).

\bibitem{tsuchiizu2002} M. Tsuchiizu {\it et. al.}, Phys. Rev. B
{\bf 66},  245106 (2002).


\bibitem{schollwock2003} U. Schollwock {\it et. al.},
Phys. Rev. Lett. {\bf 90}, 186401 (2003).


\bibitem{capponi2004} S. Capponi {\it et. al.},
Phys. Rev. B {\bf 70}, 220505(R) (2004).

\bibitem{chakravarty2001a}
S. Chakravarty, {\it et al.}, Int. J. Mod. Phys. B
{\bf  15}, 2901 (2001).


\bibitem{mook2002} H. A. Mook {\it et. al.},  \PRB ~{\bf 66},
144513 (2002).

\bibitem{mook2004} H. A. Mook {\it et. al.}, \PRB ~{\bf 69},
134509 (2004).

\bibitem{stock2002} C. Stock {\it et. al.}, \PRB~ {\bf 66},
24505 (2002).

\bibitem{murakami2003} S. Murakami, {\it et al.}, 
Science {\bf 301}, 1348 (2003).

\bibitem{sinova2004} J. Sinova {\it et. al.},
\PRL~ {\bf 92}, 126603 (2004).

\bibitem{spinhall} Y. K. Kato {\it et. al.},
Science~ {\bf 306}, 1910 (2004), J. Wunderlich {\it et. al.},
Phys. Rev. Lett. 94, 047204 (2005).


\bibitem{dzyaloshinskii-moriya} I. Dzyaloshinskii,
J. Phys. Chem. Solids {\bf 4}, 241 (1958);
T. Moriya, Phys. Rev. {\bf 120}, 91 (1960).

\bibitem{thio1988} T. Thio {\it et. al.}, \PRB~{\bf 38}, R905 (1988).

\bibitem{coffey1991} D. Coffey {\it et. al.}, \PRB~{\bf 44},
10112 (1991); D. Coffey {\it et. al.}, \PRB ~{\bf 42}, 6509 (1990).

\bibitem{bonesteel1992} N. E. Bonesteel, {\it et. al},
\PRL~ {\bf 68}, 2684 (1992).

\bibitem{shekhtman1992} L. Shekhtman et. al, \PRL~ {\bf 69}, 836 (1992);
D. Coffey et. al, \PRB~ {\bf46}, 5884 (1992);
L. Shekhtman et. al, \PRB~ {\bf 47}, 174 (1993);
W. Koshibae et al., \PRB~ {\bf 50}, 3767 (1994);
K. V. Tabunshchyk et al., \PRB~{\bf 71}, 214418 (2005) ;
M. Silva Neto et al., cond-mat/0502588.


\bibitem{FFLO}
P. Fulde {\it et. al.},, Phys. Rev. ~{\bf 135}, A550 (1964);
A. I. Larkin {\it et. al.}, Sov. Phys. JETP~ {\bf 20}, 762 (1965).


\bibitem{wu2002}
C.  Wu et. al., Phys. Rev. B {\bf 66}, 020511(R)(2002);
J.  Zhu et al., Phys. Rev. B {\bf 57}, 13410 (1998).


\bibitem{gorkov2001} L. P. Gor'kov {\it et al.}, \PRL ~{\bf 87},
37004(2001).

\end{thebibliography}
\end{document}